\documentclass[a4paper,fleqn, sort&compress]{cas-dc}

 \usepackage[numbers]{natbib}
\usepackage{graphicx}
\usepackage{url}
\usepackage{hyperref}
\usepackage{booktabs}
\usepackage{multirow}
\usepackage{subfigure}
\usepackage{amsmath}
\usepackage{amssymb}
\usepackage{color}
\usepackage{physics}
\usepackage{enumitem}
 \usepackage{mathrsfs}
\usepackage{siunitx}
\newcommand{\iu}{\mathrm{i}\mkern1mu} 
\newcommand*{\Scale}[2][4]{\scalebox{#1}{$#2$}}%

\begin{document}
\let\WriteBookmarks\relax
\def\floatpagepagefraction{1}
\def\textpagefraction{.001}

\shorttitle{A Nonlinear Damped Metamaterial: Wideband Attenuation with Nonlinear Bandgap and Modal Dissipation}

\shortauthors{Published version}

\title [mode = title]{A Nonlinear Damped Metamaterial: Wideband Attenuation with Nonlinear Bandgap and Modal Dissipation}                      



%
\author[1]{Bao Zhao}[orcid=0000-0002-9689-7742]
 \ead{bao.zhao@ibk.baug.ethz.ch}
 \cormark[1]
\credit{Conceptualization, Formal analysis, Investigation, Methodology, Resources, Writing – original draft}
\affiliation[1]{organization={Department of Civil, Environmental, and Geomatic Engineering, ETH Zürich},
    city={Zürich},
    postcode={8093}, 
    country={Switzerland}}

\author[1]{Henrik R. Thomsen}[orcid=0000-0002-8998-9091]
\credit{Validation, Investigation, Writing – review \& editing}

\author[2]{Xingbo Pu}[orcid=0000-0002-7526-9878]
\affiliation[2]{organization={Department of Mechanical and Aerospace Engineering, Hong Kong University of Science and Technology},
    addressline={Clear Water Bay, Kowloon}, 
    city={Hong Kong},
    country={China}}
\credit{Methodology, Investigation, Writing – review \& editing}

\author[3]{Shitong Fang}[orcid=0000-0002-4813-3027]
\credit{Resources, Writing – review \& editing}
\author[3]{Zhihui Lai}
\credit{Resources, Writing – review \& editing}
\affiliation[3]{organization={College of Mechatronics and Control Engineering, Shenzhen University},
    city={Shenzhen},
    postcode={518060}, 
    country={China}}

\author[4]{Bart Van Damme}[orcid=0000-0001-9289-861X]
\credit{Conceptualization, Validation, Investigation, Resources, Writing – review \& editing}
\author[4]{Andrea Bergamini}[orcid=0000-0003-2722-3207]
\credit{Resources, Writing – review \& editing}
\affiliation[4]{organization={Laboratory for Acoustics/Noise Control, Empa Materials Science and Technology},
    city={Dübendorf},
    postcode={8600}, 
    country={Switzerland}}

\author[1]{Eleni Chatzi}[orcid=0000-0002-6870-240X]
\credit{Supervision, Resources, Writing – review \& editing}

\author[5]{Andrea Colombi}[orcid=0000-0003-2480-978X]
\credit{Conceptualization, Supervision, Resources, Writing – review \& editing}
\affiliation[5]{organization={Z\"urich University of Applied Sciences (ZHAW)},
    city={Winterthur},
    postcode={8401}, 
    country={Switzerland}}

\cortext[cor1]{Corresponding author}




\begin{abstract}
In this paper, we incorporate the effect of nonlinear damping with the concept of locally resonant metamaterials to enable vibration attenuation beyond the conventional bandgap range. The proposed design combines a linear host cantilever beam and periodically distributed inertia amplifiers as nonlinear local resonators. The geometric nonlinearity induced by the inertia amplifiers causes an amplitude-dependent nonlinear damping effect. Through the implementation of both modal superposition and numerical harmonic methods with Alternating Frequency Time and numerical continuation techniques, the finite nonlinear metamaterial is accurately modeled. The resulting nonlinear frequency response reveals the bandgap is both amplitude-dependent and broadened. Furthermore, the nonlinear interaction between the local resonators and the mode shapes of the host beam is discussed, which leads to efficient modal frequency dissipation ability. The theoretical results are validated experimentally. By embedding the nonlinear damping effect into locally resonant metamaterials, wideband and shock wave attenuation of the proposed metamaterial is achieved, which opens new possibilities for versatile metamaterials beyond the conventional bandgap ranges of their linear counterparts.

\end{abstract}


\begin{highlights}
    \item A nonlinear metamaterial attenuates vibration beyond the bandgap range of the linear counterparts.
    \item Nonlinear dispersion and frequency response by homogenization and harmonic balance methods.
    \item Nonlinear damping effect enables efficient modal frequency dissipation.
    \item Experiments validate the broadening of bandgap and modal frequency dissipation.
\end{highlights}


\maketitle
\section{Introduction}

Mechanical vibrations are commonly encountered when dealing with civil infrastructures, industrial environments, vehicles, and more in general engineering applications. Their detrimental effects often lead to structural and operational failures and harm to human bodies. Therefore, vibration attenuation has received enormous attention from both research and industries. The commonly used vibration control methods include damping enhancement, stiffness tuning, and vibration mitigation by auxiliary attachments. These can be grouped into passive, semi-passive, and active methods \cite{preumont2018vibration}. Because of their easy application, passive and semi-passive methods are particularly suitable for vibration mitigation of the host structures without complex control systems. These methods can be distinguished based  on the nature of their dynamic properties as follows:
\begin{enumerate}
    \item Single linear attachment, e.g., tuned mass damper \cite{ormondroyd1928theory} and piezoelectric shunting \cite{sodano2004review};
    \item Single nonlinear attachment, e.g., nonlinear energy sinks \cite{vakakis2001inducing} and nonlinear damping \cite{starosvetsky2009vibration};
    \item Multiple linear attachments, e.g.,  multiple tuned mass dampers \cite{zuo2005optimization}, and linear locally resonant metamaterials \cite{liu2000locally};
    \item Multiple nonlinear attachments, e.g., nonlinear metamaterials \cite{casalotti2018metamaterial}.
\end{enumerate}

When it comes to vibration suppression for the host structure, either the energy is efficiently transferred to the auxiliary attachments or the energy is prevented from propagating through the host structure, thus yielding a low transmissibility on the receiver side. Conventionally, attachments could be mechanical resonators  \cite{ormondroyd1928theory,gutierrez2013tuned} or piezoelectric transducers with electrical shunting circuits \cite{sodano2004review,zhao2022circuit}. They can either transfer and dissipate the energy from the host structure or shift the resonance of the host structure by additional stiffness and mass to avoid resonances. 

However, due to their inherently linear nature, their effective bandwidth is narrow. In contrast, nonlinear attachments can interact with linear host structures in a broadband fashion. In certain conditions, this nonlinear interaction leads to a unidirectional energy flow from the linear structure to the nonlinear attachments. In the context of nonlinear energy sinks \cite{vakakis2008nonlinear}, this is referred to as nonlinear energy transfer. In addition, the nonlinear reaction forces induced by nonlinear stiffness \cite{zhao2020dual,zhao2023jump}, damping \cite{starosvetsky2009vibration}, or vibro-impact \cite{gzal2021extreme} essentially couple the separated modes of the linear host structure and result in energy transfer, redistribution, and efficient dissipation among different structural modes \cite{gzal2020rapid}. 

When moving from single attachments to multiple attachments, the challenge lies in how to design and optimize multiple attachments or resonators to effectively attenuate the vibration of the host structure, which normally has multiple vibration modes. From a modal analysis point of view, attenuation at modal frequencies forms the guideline for designing different tuned mass dampers \cite{zuo2005optimization}. From a wave propagation perspective, vibration modes are attributed to reflections from the domain boundaries. Therefore, to effectively suppress vibration modes, the propagation of the traveling wave from one end of the host structure to the other must be prohibited for certain frequency ranges. This is synonymous with the design of bandgaps in locally resonant metamaterials \cite{liu2000locally}. Through the proper tuning of mechanical parameters, the propagating wave can be gradually trapped \cite{colombi2016seismic,davies2023graded}, redirected \cite{pu2022topological, de2023tailored,thomsen2023boundless}, or absorbed \cite{zhao2022graded,rosafalco2023optimised} with periodically distributed local resonators attached on the host structure, resulting in zero group velocities and low transmissions.

The recent endeavors to introduce nonlinearities through multiple nonlinear attachments have combined the advantages of locally resonant metamaterials and nonlinear dynamics. This has stimulated the emergence of novel concepts in the context of structural dynamics, such as harmonic generations \cite{fang2022nonlinear}, chaotic bands \cite{fang2017ultra}, and broadband vibration attenuation \cite{xia2020bistable,cho2022computational}.  However, owing to the challenges of relatively high degrees of freedom in metamaterials and complex dynamics from nonlinear attachments, theoretical and experimental realizations in this area have been incomplete until recent years. From a theoretical perspective, the high number of degrees of freedom, different nonlinear forms, and intensities of nonlinearities pose difficulties to reach an accurate solution. Compared to the numerical integration method, conventional analytical low-order methods such as harmonic balance methods \cite{fang2017ultra,fang2022nonlinear,xia2020bistable,xu2023vibration} and perturbation methods \cite{bae2022nonlinear,bae2020amplitude} fail to converge when the nonlinearities presented are strong, even without accounting branch bifurcations due to stability issues \cite{seydel2009practical}. In addition, these analytical methods, which work for close-form polynomial nonlinearities such as cubic stiffness \cite{fang2017ultra,xia2020bistable} and quadratic damping \cite{xu2023vibration}, are often insufficient for more general nonlinearities such as non-smooth nonlinear damping studied in this paper and piece-wise nonlinear forces.  Theoretical advances are further complicated because standard Bloch-Floquet theory is limited in its usability. This popular method cannot reveal the modal interactions of multiple nonlinear attachments in a finite metamaterial, or frequency coupling related to broadband excitation. From an experimental perspective, nonlinear metamaterials need to be designed, fabricated, and tested from the resonator level to the metamaterial level to confirm the proposed nonlinear effect in practical manners, which requires repeatable and slow sampling with the increase deviation of the nonlinearity of the system from its linear state \cite{fang2017ultra,fang2022nonlinear,xia2020bistable,cho2022computational}. In addition, the amplitude-dependent response and the nonlinear coupling in nonlinear metamaterials also open new possibilities that need to be studied for vibration attenuation or energy exchange beyond the frequency range of the linear bandgap \cite{fang2017ultra,fang2022nonlinear}. From an ideology perspective, the effects of general nonlinearities in conventional low degree of freedom systems such as nonlinear vibration absorbers \cite{vakakis2001inducing} have been well studied and documented. Their practical applications and accurate modeling methods in nonlinear metamaterials \cite{fang2022nonlinear,cho2022computational}, rather than conceptual verification of the known effects, are of great importance toward broader studies and engineering applications.

Based on our previous published  conference paper \cite{Zhao2023patra}, we present modeling methods for Euler-Bernoulli beam-based nonlinear metamaterials with general local nonlinearities and investigate a novel nonlinear damped metamaterial for wideband vibration attenuation and modal dissipation with a practical design of the nonlinear local resonators. With the goal of designing and solving real nonlinear metamaterial systems, we establish all the analyses on practical experimental parameters.  We utilize a dispersion analysis and a modal analysis method with a numerical harmonic balance method to solve the amplitude-dependent response of this nonlinear metamaterial. Through the Alternating Frequency Time (AFT) and numerical continuation techniques, we can handle more general nonlinearities with the possibility of harmonic generations for weak to strong nonlinear scenarios. In addition, the nonlinear modal coupling is discussed to demonstrate the nonlinear interaction between the local resonators and the mode shapes of the host beam, which leads to efficient modal energy dissipation ability beyond the study of the conventional bandgaps. Based on the theoretical analysis, a nonlinear metamaterial is designed to incorporate the geometric nonlinear damping effect induced by the inertia amplifiers. The nonlinear frequency response of a single nonlinear resonator and the transmissibility of the nonlinear metamaterial are measured experimentally. The theoretical and experimental results not only validate the methods for solving the amplitude-dependent responses of the proposed nonlinear metamaterial but also give insights into the mechanisms for wideband attenuation combining nonlinear bandgap and modal frequency dissipation, which opens new possibilities for versatile metamaterials surpassing the limit of their linear bandgap ranges.


\section{Theoretical Analysis}
\label{sec:theor}
\begin{figure}[!t]
\centering
\includegraphics[width=1\columnwidth,page=1]{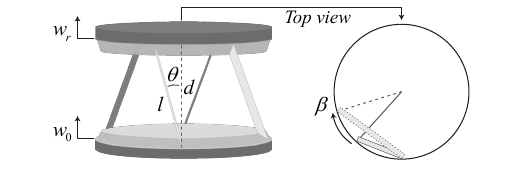}
\caption{The schematic and top view of a rotational inertia amplifier.}
\label{fig:model}
\end{figure}

\begin{figure}[!t]
\centering
\includegraphics[width=1\columnwidth,page=2]{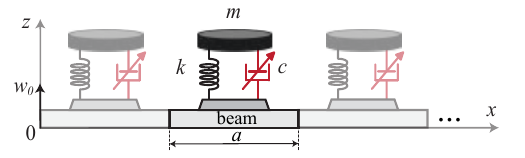}
\caption{A semi-infinite schematic of the nonlinear metamaterial with lattice constant $a$. The incident wave $w_0$ in $z$ direction is applied at the origin $x=0$.}
\label{fig:nm_1}
\end{figure}

 The proposed nonlinear metamaterial is shown in Fig. \ref{fig:nm_1}, which illustrates a semi-infinite case in the $x$-direction of the nonlinear metamaterial consisting of a host beam and inertia amplifiers as nonlinear local resonators. Before analyzing the nonlinear metamaterial, we first revisit the nonlinear dynamics of the rotational inertia amplifier proposed by Van Damme et al. \cite{van2021inherent}. As shown in Fig. \ref{fig:model}, the inertia amplifier consists of two identical disks with mass $m_0$ on the top and bottom and four beams as connections that are tilted with angle $\theta$. The tilted beams lead to a coupled translation-rotation motion of the top disk and simultaneously deliver a longitudinal spring stiffness $k_0$. The rotational spring stiffness is considered to be small compared to the longitudinal spring stiffness, which can be ensured by sufficiently thin connector points.

The nonlinearity is induced by the coupling of the translational motion of the bottom and the rotational motion of the top disk. The relative displacement of the top disk with respect to the bottom disk is denoted as $w_r$. And the rotation angle of the top disk is denoted as $\beta$.
In Fig. \ref{fig:model}, the top view shows the projected deformation of a connected beam, while the chord length in the top view of Fig. \ref{fig:model} changes from $\sqrt{l^2-d^2}$ to $\sqrt{l^2-(d-w_r)^2}$. For small $w_r$, the top disk's rotation angle $\beta$ is defined as:
\begin{equation}
\beta=\frac{\sqrt{l^2-(d-w_r)^2}-\sqrt{l^2-d^2}}{R}
\label{eq:beta}
\end{equation}
where $R$ is the radius of the disk. The angular velocity can be written as:
\begin{equation}
\dot{\beta}=\frac{\dot{w}_r}{R} \frac{d-w_r}{\sqrt{l^2-(d-w_r)^2}} .
\label{eq:vbeta}
\end{equation}
where $l=d/\cos\theta$. The kinetic energy of the top disk can be defined as $T=m_0|\dot{w}_r|^2/2+I_0|\dot{\beta}|^2/2$, where $I_0=m_0R^2/2$ represents the moment of inertia of the rotating disk. Its potential energy is $V=k_0 w^2_r/2$. Using the Lagrangian $\mathscr{L}=T-V$, its equation of motion can be formulated as:
\begin{equation}
\begin{aligned}
    F=-m\ddot{w}_0&=\frac{d}{dt}\left( \frac{\partial \mathscr{L}}{\partial \dot{w}_r }\right)-\frac{\partial \mathscr{L}}{\partial{w}_r } \\
    &={m}\ddot{w}_r+{c}|\dot{w}_r|\dot{w}_r+{k}w_r,
\end{aligned}
\label{eq:eom_res}
\end{equation}
where $F=-m\ddot{w}_0$ is the harmonic base excitation force applied on the bottom disk with the displacement and displacement amplitude denoted as $w_0$ and $W_0$, respectively. By substituting Eq. \ref{eq:vbeta} into Eq. \ref{eq:eom_res}, the equivalent mass $m$, damping $c$, and stiffness $k$ can be determined as: 
\begin{equation}
\begin{aligned}
m & =m_0+\frac{m_0}{2}\left(1+\frac{(1-\varepsilon)^2}{A}\right) \\
c & =\frac{m_0}{2 d}\left(\frac{1-\varepsilon}{A}+\frac{(1-\varepsilon)^3}{A^2}\right) \\
k & =k_0 \\
A & =\frac{1}{\cos ^2 \theta}-(1-\varepsilon)^2,
\end{aligned}
\label{para_res1}
\end{equation}
where $\varepsilon=w_r/d$ represents the strain along the translational direction. Under small strain $\varepsilon$ condition, the Taylor series of the velocity $\dot{\beta}$ reads: 
\begin{equation}
  \dot{\beta}\left(\varepsilon\right)=\frac{\dot{w}_r}{R}\left[ \frac{1}{\tan\theta}+\frac{\varepsilon}{\tan\theta \sin^2\theta}+\frac{3}{2}\frac{\varepsilon^2}{\tan^3\theta \sin^2\theta } \right].
  \label{eq:talyor}
\end{equation}
Due to the nonlinear geometric constraint between displacement and rotation angle, $\dot{\beta}\left(\varepsilon\right)$ not only linearly depends on $\dot{w}_r$ with the $0$-th order term, the second, third, and high order terms with respect to $\varepsilon$ also play important roles under sufficiently small $\theta$ cases. However, by assuming the strain $\varepsilon$ no more than 2.5\%, we can calculate the smallest $\theta=\ang{30}$ that makes the second, third, and higher order terms 10 times smaller than the $0$-th order coefficient in Eq. \ref{eq:talyor}.

In this case, the $0$-th order term with respect to $\varepsilon$ of Eq. \ref{eq:talyor}, together with the small strain $\varepsilon$ condition, can be used to simplify the equivalent parameters in Eq. \ref{para_res1}:
\begin{equation}
  m=m_0+\frac{m_0}{2\sin^2\theta}, \quad  c=\frac{m_0}{2d}\frac{\cos^2\theta}{\sin^4\theta}, \quad k=k_0,
  \label{eq:para_res}
\end{equation}
where $k$ is the spring constant depending on the connections between two disks. From the expression of $m$, the dynamically added mass due to the top disk is $m_0/2\sin^2\theta$ with the amplification factor $\alpha=1/\sin^2\theta$. A nonlinear damping effect emerges from the expression of $c$ due to the geometric nonlinear coupling between $\dot{w}_r$ and $\dot{\beta}$. This nonlinear damping force ${c}|\dot{w}_r|\dot{w}_r$, also known as the drag force \cite{worden2001nonlinearity}, has an absolute form with respect to the velocity, which can be cataloged into non-smooth nonlinearities. This enables the amplitude-dependent frequency response and amplitude-dependent bandgaps in nonlinear metamaterials \cite{bae2020amplitude}. 

By assuming a fundamental harmonic solution for $w_r=W_r\sin(wt)$ with ${W}_r$ representing the displacement amplitude, we can determine a nonlinear correspondence to the linear viscous damping coefficient, an equivalent damping coefficient $c_{eq}={c}|\dot{w}_r|$ that increases with the base excitation force. By treating the absolute value with sign function, Eq. \ref{eq:eom_res} can be written as:
\begin{equation}
{m}\ddot{w}_r+{c}\cdot\text{sign}(\dot{w}_r)\dot{w}_r^2+{k}w_r=-m\ddot{w}_0,
\label{eq:eom_res1}
\end{equation}
where $\text{sign}(\dot{w}_r)$ is an even square wave with its leading harmonic $4\cos(\omega t)/\pi $. Substituting the solutions into Eq. \ref{eq:eom_res1}, and balancing the first order harmonic, yields:
\begin{equation}
\left(\left(k-m \omega^2 \right)^2+ \left( 3c\omega {W}_r/\pi \cdot \omega\right)^2 \right){W}_r^2=\left(\omega^2mW_0\right)^2,
\label{eq:frf1}
\end{equation}
Observing the above equation, the equivalent nonlinear damping reads:
\begin{equation}
    c_{eq}=3c\omega{W}_r/\pi,
    \label{eq:ceq}
\end{equation}
where ${W}_r$ is given as:
\begin{equation}
{W}_r=\frac{\omega^2mW_0}{\sqrt{(k-m \omega^2)^2+9 c^2 \omega^4 {W}_r^2/\pi^2}}.
\end{equation}

It can be seen that $c_{eq}$ is a function of frequency $\omega$ and the base excitation amplitude $W_0$. To illustrate the nonlinear effects of the described nonlinear resonator in the proposed locally resonant metamaterial, we use the practical parameters in Table \ref{para} for the analyses in the following subsections. The frequency ranges presented in the following figures are normalized with respect to the linear resonant frequency of the inertia amplifiers at 210.7 Hz.

\begin{figure*}
\centering
\includegraphics[width=2\columnwidth,page=3]{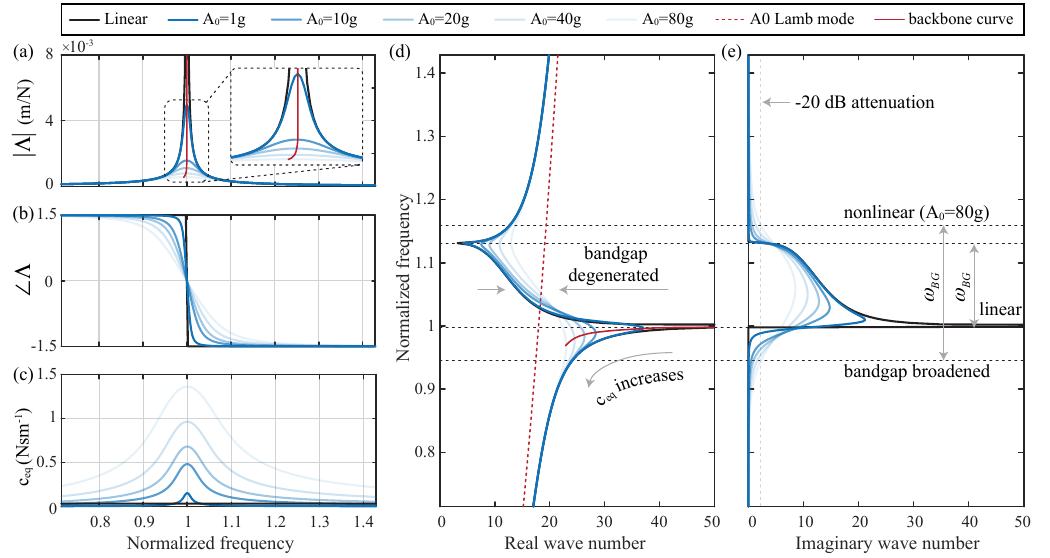}
\caption{The frequency response function, equivalent damping, and dispersion curves under different acceleration amplitudes. (a) and (b): The amplitude and phase of the frequency response function of a local resonator. The enlarged view in (a) shows the shift of the resonant frequency indicated with a red backbone curve; (c) The equivalent damping of a local resonator; (d) and (e): The real and imaginary wave number of the nonlinear metamaterial.}
\label{fig:disper}
\end{figure*}

\subsection{Nonlinear Dispersion Relationship}
\label{subsec:dispersion}
The dispersion relationship provides a general and fundamental description of the wave propagation characteristics of metamaterials. Unlike linear metamaterials, nonlinear metamaterials give rise to amplitude-dependent dispersion relationships \cite{sepehri2022wave,lou2022propagation}, which leads to a better understanding of how the nonlinearity could enable rich dynamics in metamaterials.

We herein discuss the dispersion relationship for flexural waves traveling in the proposed nonlinear metamaterial, where the nonlinearity stems from the resonator's amplitude-dependent damping. This mechanism is much less investigated than amplitude-dependent stiffness changes. As shown in Fig. \ref{fig:nm_1}, the lattice constant of $j$-th unit cell is $a$. The linear density of the host beam is $\rho_0=\rho bh$, where $\rho$, $b$, and $h$ represent the density, width, and height of the host beam, respectively. The bottom disk with mass $m_0$ of the inertia amplifier is fixed on the beam. It is coupled with an equivalent mass $m$ by a linear spring $k$ and a nonlinear damper $c$. An incident wave $w_0=W_0\sin(\omega t+\varphi)$ in $z$ direction is applied at the boundary $x=0$ of the beam, where $W_0$ is the amplitude of the incident wave. Rather than using a complex-valued frequency and real-valued wavenumber \cite{hussein2009theory}, we adopt a real-valued frequency and allow for complex-valued wavenumbers, which is suitable for forced harmonic cases \cite{palermo2022rayleigh,hussein2010band} and gives insight in the attenuation of propagating waves \cite{van2017impact,van2018measuring}.

With reference to the parameters from Table \ref{para}, we can obtain the dispersion curve of the A0 Lamb mode of the host beam by Euler-Bernoulli beam theory as shown with the red dash line in Fig. \ref{fig:disper}(d). The flexural wavelength of the host beam is inversely proportional to the wave number with $2\pi$, which is approximately 35 cm at the linear resonant frequency of the inertia amplifiers. Since the lattice constant is sufficiently smaller than the wavelength under low-frequency vibrations around this frequency range, the wave profile between two adjacent unit cells can be approximated by a smooth function by neglecting the near-field scattering around the inertia amplifiers. Thus, we can utilize the averaging technique from the homogenization method \cite{lou2022propagation} and transform the concentrated reaction force of the local resonator to a uniformly distributed force $f(x,t)$ applied evenly with lattice constant $a$.  The governing equations of the nonlinear metamaterial are given as:
\begin{equation}
\begin{aligned}
      &D_0 \frac{\partial^{4} w\left(x,t\right)}{\partial x^{4}}+\rho_0 \frac{\partial^{2} w \left(x,t\right)}{\partial t^{2}}=-f(x,t)\\
      &m(\ddot{w}_r+\ddot{w})=-k w_r-c |\dot{w}_r|\dot{w}_r\\
      &m_0\ddot{w}+m(\ddot{w}_r+\ddot{w})=af(x,t)
\end{aligned},
  \label{eq:homo}
\end{equation}
where $D_0=EI$ is the flexural rigidity of the host beam. $w$ is the transverse displacement at beam position $x$. And $w_{r}$ is the relative displacement of the local resonator at the same beam position. By adopting the effective medium theory \cite{choy2015effective}, Eq. \ref{eq:homo} can be transferred into:
\begin{equation}
\left\{\begin{aligned}
&D_0\frac{\partial^4 w}{\partial x^4}+(\rho_0+\frac{m_0+m}{a}) \frac{\partial^2 w}{\partial t^2}+\frac{m}{a} \frac{\partial^2 w_r}{\partial t^2}=0 \\
&m \frac{\partial^2}{\partial t^2}(w_r+w)+k w_r+c |\dot{w}_r|\dot{w}_r=0
\end{aligned}. \right.
  \label{eq:homo2}
\end{equation}
By  neglecting the higher harmonic generations, we can replace the nonlinear damping force $c|\dot{w}_r|\dot{w}_r$ with an equivalent damping force $c_{eq}\dot{w}_r$ and only consider the fundamental harmonic traveling wave $w=We^{\iu(\omega t-kx)}$. The dispersion relationship for the metamaterial beam can be derived from Eq. \ref{eq:homo2} as:
\begin{equation}
k(\omega)=\left[\left(\rho_0 +\frac{m_0}{a}+\frac{k+\iu c_{eq}\omega}{a(\omega_r^2-\omega^2+\iu c_{eq}\omega/m)}\right) \frac{\omega^2}{D_0}\right]^{\frac{1}{4}},
\label{eq:dispersion}
\end{equation}
where $\omega_r$ is the linear resonant frequency of the inertia amplifier. Therefore, the effective mass density for the proposed metamaterial beam reads:
\begin{equation}
\rho_{e} = \frac{1}{a}\left( \rho_0a +m_0+ \frac{k+\iu c_{eq}\omega}{\omega_r^2-\omega^2+\iu c_{eq}\omega/m} \right).
\label{eq:density}
\end{equation}
If $c_{eq}=0$, the effective mass density and the dispersion for a metamaterial beam recover to its classical linear and undamped case \cite{sugino2016mechanism}; if $c_{eq}=\text{constant}$, then it represents a linear damped case \cite{palermo2022rayleigh}; if $c_{eq}(\omega,W_0)$ is a function of frequency $\omega$ and excitation amplitude $W_0$ as shown in Eq. \ref{eq:ceq}, then this indicates a nonlinear damping case induced by the inertia amplifiers. 

From a practical perspective, we prescribe a certain acceleration amplitude $A_0$ as excitation. Therefore, the equivalent damping $c_{eq}$ is frequency and acceleration dependent. By solving Eq. \ref{eq:frf1}, an equivalent local resonator with \textit{Frequency Response Function} $\Lambda$ can be achieved, together with Eq. \ref{eq:dispersion}. The results are shown in Fig. \ref{fig:disper} under different acceleration amplitudes. The effect of the increasing damping can clearly be seen by a reduction of the resonator's amplitude and a shift to lower eigenfrequencies (Fig. \ref{fig:disper} (a)), as well as the decrease of the phase slope (Fig. \ref{fig:disper} (b)). 

By prescribing a sufficiently small $A_0$, the results of a linear metamaterial beam can be recovered as black curves in Fig. \ref{fig:disper}. With the increase of $A_0$, the resonant frequency slightly shifts to a lower frequency due to the increase of equivalent damping $c_{eq}$ as shown by the backbone curve in Fig. \ref{fig:disper} (a). In Fig. \ref{fig:disper} (c),  the equivalent damping ($c_{eq}$) reaches the maximum at the linear resonant frequency $\omega_r$. 


For undamped linear metamaterials, $\rho_e$ is real-valued. The bandgap is formed when $\rho_e<0$, and hence the wave number $k(\omega)$ is complex-valued. This convenient criterion gives the classical bandgap range \cite{sugino2016mechanism}. However, when it comes to damped metamaterials, $\rho_e$ and  $k(\omega)$ are always complex-valued, making the bandgap definition unpractical. In this paper, we define the bandgap range as the frequency range in which the transmissibility of a traveling wave along a unit length is reduced by more than -20 dB, which is commonly used as a threshold for vibration attenuation. Therefore, with the assumption of the fundamental harmonic traveling wave, the bandgap range $\omega_{BG}$ reads:
\begin{equation}
\left\{\omega_{BG} \in \mathbb{R} \mid 20\log_{10}e^{-|\Im(k(\omega))|}<-20\right\}.
\label{eq:bandrange}
\end{equation}

The dispersion curves for the proposed nonlinear metamaterial are shown in Fig. \ref{fig:disper}(d) and (e). With the increase of $A_0$, the real and imaginary parts of $k(\omega)$ become smoother than in the linear case. This is caused by the gradual increase of the damping effect from the local resonators, as shown in Fig. \ref{fig:disper}(c). For the real part of $k(\omega)$ in Fig. \ref{fig:disper}(d), the real wave number becomes smaller with the increase of $A_0$, which bends its in-phase branches from an infinite value to a finite value and thus leads to less attenuation. This degeneration of the dispersion curve also gives rise to the partial wave number bandgap beyond which the wave propagation is forbidden \cite{farzbod2011analysis,hussein2010band}. The spatial decay of the traveling wave can be observed from the imaginary part of $k(\omega)$ in Fig. \ref{fig:disper}(e). An increase of $A_0$ not only reduces the resonant frequencies of local resonators for lower beginning frequencies of bandgaps but broadens their bandwidth, which consequentially broadens the bandgap range of the nonlinear metamaterial. Due to the nonlinear damping, the nonlinear bandgap range is broader than its corresponding linear case. It should be noted that this broadening effect comes  with a slight bandgap degeneration. Further increase of $A_0$ could reversely reduce the bandgap range if a certain attenuation level is desired. 

\subsection{Nonlinear Frequency Response}
\label{subsec:frf}

\begin{figure*}
\centering
\includegraphics[width=2\columnwidth,page=4]{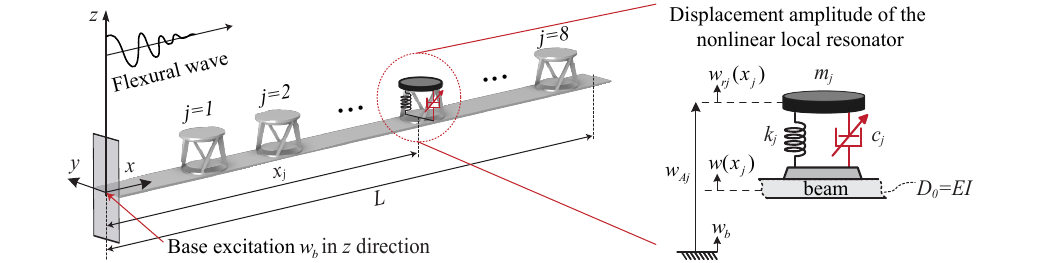}
\caption{The model of the nonlinear metamaterial with finite length and clamped-free boundary condition. The enlarged view shows the displacement components for the absolute transverse displacement amplitude $w_{Aj}$ of the nonlinear local resonator at $x_j$ along the host beam. }
\label{fig:nm_2}
\end{figure*}

Unlike the nonlinear dispersion relationship mentioned above, the infinite long beam assumption practically does not hold due to boundary reflections under low-frequency vibrations. Modal frequencies of the entire structure induced by boundary conditions interact with the nonlinear local resonators.
Therefore, in this part, we discuss the frequency response of the proposed nonlinear metamaterial with a finite length by modal analysis and consider the higher harmonics through the harmonic balance method.

As shown in Fig. \ref{fig:nm_2}, we consider a clamped-free cantilever beam with length $L=0.6$ m and in total $S=8$ inertia amplifiers periodically distributed along the beam from $x_1=0.14$ m to $x_9=0.56$ m. In this paper, we focus on the out-of-plane flexural wave propagation rather than the in-plane wave propagation of the host beam. In order to suppress the undesired in-plane modes, the inertia amplifiers are designed with four beams as connections to the bottom disk attached on the host beam to symmetrically distribute the reaction forces. However, mode conversion could happen at the crossings of dispersion curves when asymmetry or chirality are involved with three beams as connections \cite{bergamini2019tacticity}. The detailed parameters of the cantilever beam are listed in Table. \ref{para}. The absolute transverse displacement, $w_{Aj}\left(t\right)$, of the $j$-th nonlinear resonator at $x_j$ along the host beam is defined as:
\begin{equation}
w_{Aj}\left(t\right)=w_b\left(t\right)+w\left(x_j,t\right)+w_{rj}\left(t\right),
\end{equation}
where $w_b\left(t\right)$ is the base excitation displacement at the clamp side and $w_{rj}\left(t\right)$ is the relative displacement of $j$-the nonlinear resonator. The equation of motion of each nonlinear local resonator can be formulated as follows:
\begin{equation}
    {m_j}\left( {\frac{{{\partial ^2}w}}{{\partial {t^2}}} + {\ddot w}_b}+{\ddot w}_{rj} \right) + {c_j}|{\dot w}_{rj}|{\dot w}_{rj} + {k_j}w_{rj}  =  0,
    \label{eq:eom_res2}
\end{equation}
where $m_j$, $c_j$ and $k_j$ are determined by Eq. \ref{eq:para_res}. By adding the reaction forces of each nonlinear resonator onto the host beam, the governing equation of the host beam can be expressed as:
\begin{equation}
\begin{aligned}
&D_0 \frac{\partial^{4} w}{\partial x^{4}}+\rho_0 \frac{\partial^{2} w}{\partial t^{2}}=-\rho_0 \ddot{w}_{b}+ \\
&\sum_{j=1}^{S} \left(k_{j} w_{rj}+ {c_j}|{\dot w}_{rj}|{\dot w}_{rj}\right) \delta\left(x-x_{j}\right),
\end{aligned}
\label{eq:govern}
\end{equation}
where $\delta$ represents the Dirac function. We here modify $\rho_0=\rho bh+m_0/a$ to take into account the inertia force of the bottom disks. The nonlinear metamaterial is then represented by Eq. \ref{eq:eom_res2} and  Eq. \ref{eq:govern}.

Assume the total mass of local resonators is much smaller than the mass of the host beam and linear vibration modes dominate the response of the system so that the nonlinearities induced by the nonlinear damping effect can be treated as perturbations to the underlying linear Euler-Bernoulli equation of host beam \cite{brunton2022data,xia2020bistable}. Compared with the conventional finite element method (FEM), the modal superposition method takes advantage of reduced order modeling, effectively reducing the problem's dimension \cite{rouleau2017comparison}. Thus, the approximated solution of the displacement of the host beam is found by combining the harmonic balance and modal superposition methods, defined as:
\begin{eqnarray}
w\left(x,t\right)=\sum_{i=1}^N \eta_i(t)\phi_i(x) ,
\label{eq:w}
\\
\eta_i(t)=\sum_{h=-H}^H \hat{\eta}_i(h)e^{\iu h\omega t},
\label{eq:eta_r}
\end{eqnarray}
where $N$ and $H$ are the number of modes and harmonics order considered. $\phi_i(x)$ is the $i$-th order mode shape of the host beam with its natural frequency $\omega_i$. The modal weight $\eta_i(t)$ is expanded as Fourier series for higher harmonic generations due to the nonlinear reaction forces in the system, where $\hat{\square}$ represents the complex Fourier coefficient. 

\begin{figure*}
\centering
\includegraphics[width=2\columnwidth,page=5]{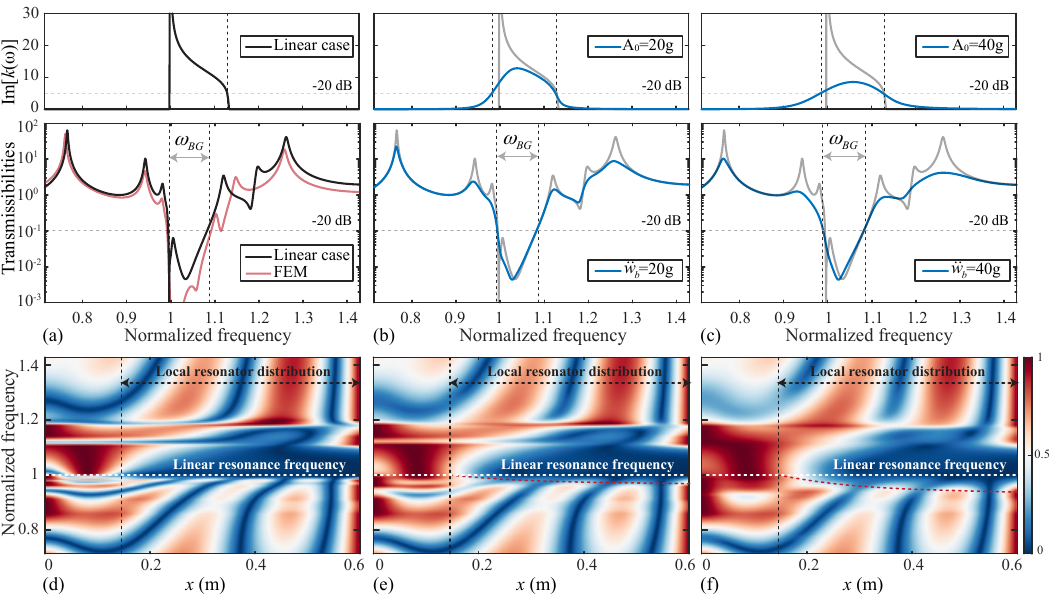}
\caption{The tip transmissibilities and spatial frequency analyses of the nonlinear metamaterial. (a)-(c): Tip transmissibilities for linear, $\ddot{w}_b=20$g nonlinear, and $\ddot{w}_b=40$g nonlinear cases; (d)-(f): Spatial frequency analyses demonstrate the normalized displacement amplitude along the host beam for linear, $\ddot{w}_b=20$g nonlinear, and $\ddot{w}_b=40$g nonlinear cases.}
\label{fig:fre}
\end{figure*}

By applying the boundary conditions of the clamped-free host beam, substituting the ansatz of the solutions, and applying orthogonality \cite{sugino2016mechanism,xia2020bistable} (detailed in Appendix A), the second-order differential equations  Eq. \ref{eq:eom_res2} and  Eq. \ref{eq:govern} can be written into matrix form with the dimension of $N+S$ as:
\begin{equation}
\mathbf{M\ddot u + Ku+ F_{nl}(\dot{u})=F_{ex}(t)},
\label{eq:matrix}
\end{equation}
where $\mathbf{u}=\left[\begin{array}{l}\eta_{1} \eta_{2} \cdots \eta_{N} w_{r1} w_{r2} \cdots w_{rS}\end{array}\right]^\intercal$ describes the modal weights and relative displacements of the nonlinear local resonators. 

For the linear part of Eq. \ref{eq:matrix}, the detailed form of the mass matrix $\mathbf{M}$ and the linear stiffness matrix $\mathbf{K}$  are given as:
\begin{equation}
\mathbf{M}=\left[\begin{array}{cc}
\mathbf{M}_{11} & \mathbf{M}_{12} \\
\mathbf{M}_{21} & \mathbf{M}_{22}
\end{array}\right],
\quad
\mathbf{K}=\left[\begin{array}{cc}
\mathbf{K}_{11} & \mathbf{0} \\
\mathbf{0} & \mathbf{K}_{22}
\end{array}\right],
\label{eq:MK}
\end{equation}
where $\mathbf{M}_{11}$ is a $N\times N$ matrix with the entries: $m_{mn}=\delta_{mn}+\sum_{j=1}^S m_j\phi_m(x_j)\phi_n(x_j)$; $\mathbf{M}_{12}$ is a $N\times S$ matrix with the entries: $m_{mq}=m_q\phi_m(x_q)$; $\mathbf{M}_{21}$ is a $S \times N $ matrix with the entries: $m_{pn}=m_p\phi_n(x_p)$; $\mathbf{M}_{22}$ is a $S \times S $ matrix with the entries: $m_{pq}=\delta_{pq}m_p$; $\mathbf{K}_{11}$ is a $N \times N $ diagonal matrix with the entries: $k_{mn}=\delta_{mn}\omega_m^2$; $\mathbf{K}_{22}$ is a $S \times S $ diagonal matrix with the entries: $k_{pq}=\delta_{pq}k_{rp}$; The applied external force $\mathbf{F_{ex}}$ with $N+S$  items can be expanded as:
\begin{equation}
\begin{aligned}
    \mathbf{F_{ex}}=[& q_{1}\cdots q_{i}\cdots q_{N}\\
    & -m_1\ddot{w}_b\cdots -m_j\ddot{w}_b\cdots -m_S\ddot{w}_b]^\intercal,
\end{aligned}
\label{eq:fex}
\end{equation}
where $q_i$ is the modal force defined in Appendix A. 

For the nonlinear part, the nonlinear force $\mathbf{F_{nl}}(\dot{u})$ can be expressed as:
\begin{equation}
  \mathbf{F_{nl}}(\dot{u})= \mathbf{C_N|\dot u|\dot u}, \quad \mathbf{C_N}=\left[\begin{array}{cc}
  \mathbf{0} & \mathbf{0} \\
  \mathbf{0} & \mathbf{C}_{22}
  \end{array}\right],
  \label{eq:fnl}
\end{equation}
where $\mathbf{C_{N}}$ represents the nonlinear damping coefficient matrix. $\mathbf{C}_{22}$ is a $S \times S $ diagonal matrix with the entries: $C_{pq}=\delta_{pq}c_{rp}$.

By treating the relative displacements of the nonlinear resonators also as Fourier series and substituting the ansatz of $\mathbf{u}(t)$: $\mathbf{u}(\hat{\mathbf{u}},t)$ into Eq. \ref{eq:matrix}, we can form the residual function $\hat{\mathbf{r}}(\hat{\mathbf{u}}, \omega)$ by using the harmonic balance method up to the truncation order $H$:
\begin{equation}
\begin{aligned}
\hat{\mathbf{r}}(\hat{\mathbf{u}}, \omega)&=\left(\nabla^2 \otimes \omega^2 \mathbf{M} + \nabla^0 \otimes \mathbf{K}\right) \hat{\mathbf{u}}\\
&+\hat{\mathbf{F}}_{\mathrm{nl}}(\hat{\mathbf{u}}, \omega)-\hat{\mathbf{F}}_{\mathrm{ex}}(\omega)=\mathbf{0},
\end{aligned}
\label{eq:hbres}
\end{equation}
where $\nabla=\mathrm{diag}[-\mathrm{i}H,\ldots,\mathrm{i}H]$ is a diagonal matrix of dimension $2H+1$. Eq. \ref{eq:hbres} shows that the linear internal and external forces are decoupled for different harmonic indices $k$ except for the nonlinear forces $\hat{\mathbf{F}}_{\mathrm{nl}}$. With $M$, $K$, and  $\hat{\mathbf{F}}_{\mathrm{ex}}$ given, the linear parts of Eq. \ref{eq:hbres} can be easily solved, which represents the conventional linear metamaterials setup \cite{sugino2016mechanism}. The primary challenge is to determine the Fourier coefficients of the nonlinear forces.

Since $\mathbf{F_{nl}}$ is a $\mathcal{C}^1$ continuous nonlinear function without a high degree of smoothness, conventionally, it requires tedious expansions of closed-form expression up to high truncation orders in the frequency domain for convergence. While in the time domain, the nonlinear forces are easy to calculate with the available state histories of the system. By taking the advantages of the two domains, Alternating Frequency–Time (AFT) \cite{Cameron1989} resolves the Fourier coefficients of the nonlinear force by performing Fourier transforms in the frequency domain with the nonlinear forces evaluated in the time domain as:
\begin{equation}
  \hat{\mathbf{F}}_{\mathrm{nl}}\approx \hat{\mathbf{F}}_{\mathrm{nl}}^{\mathrm{AFT}}=\mathscr{F}\left[ \mathbf{F}_{\mathrm{nl}}\left(\mathscr{F}^{-1}\left[\omega\nabla\hat{\mathbf{u}}\right]\right)\right],
  \label{eq:21}
\end{equation}
where $\mathscr{F}$ denotes the discrete Fourier transform. By taking sufficient sampling points, the inverse Fourier transform $\mathscr
{F}^{-1}\left[\omega\nabla\hat{\mathbf{u}}\right]$ gives the generalized velocities in the time domain, which are used to generate the nonlinear forces at the sampling instants. Finally, the discrete Fourier transform approximates the Fourier coefficients for the nonlinear force.

Based on the work by Krack and Gross \cite{krack2019harmonic}, the nonlinear frequency response of the proposed nonlinear metamaterial within the frequency range $\left[\omega_s,\omega_e\right]$ can be solved by balancing the residual function Eq. \ref{eq:hbres} with unknown Fourier coefficients through a numerical Newton method. Compared with the direct time domain integration method, the harmonic balance method has an outstanding convergence rate due to the prior periodic ansatz of solutions. In addition, the harmonic balance method does not have the transient evolution of the time domain integration. Thus, it does not need a criterion for periodic behavior that stands for the steady state.
Nevertheless, Eq. \ref{eq:matrix} can still be solved by the time-domain Runge-Kutta method as a reference with the state-space form:
\begin{equation}
\dot{\mathbf{z}}=\mathbf{B}\mathbf{z}+\mathbf{C |\dot z|\dot z}+\mathbf{D},
\label{eq:td}
\end{equation}
where
\begin{equation}
\begin{aligned}
\mathbf{z}&=\left[\begin{array}{c}
\mathbf{u} \\
\dot{\mathbf{u}}
\end{array}\right],
&&\mathbf{B}=\left[\begin{array}{cc}
\mathbf{0} & \mathbf{I} \\
-\mathbf{M^{-1}K} & \mathbf{0}
\end{array}\right], \\
\mathbf{C}&=\left[\begin{array}{cc}
\mathbf{0} & \mathbf{0} \\
\mathbf{0} & -\mathbf{M^{-1}C_N}
\end{array}\right],
&&\mathbf{D}=\left[\begin{array}{c}
\mathbf{0} \\
\mathbf{M^{-1}F_{ex}}
\end{array}\right].
\end{aligned}
\label{eq:23}
\end{equation}

\begin{figure}
\centering
\includegraphics[width=1\columnwidth,page=6]{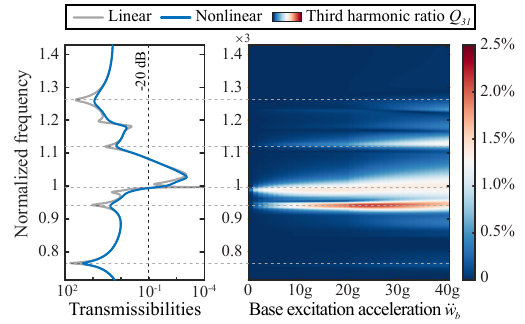}
\caption{Third harmonic generation ratio $Q_{31}$ between the displacement amplitude of the third harmonic at the free tip of the host beam and that of the first order harmonic. The gray dashed lines correspond to the third harmonic generation at modal frequencies.}
\label{fig:3rd}
\end{figure}

It should be noted that the positions of local resonators could influence the condition number \cite{golub2013matrix} of matrix $\mathbf{M}$ in Eq. \ref{eq:MK}, which measures how sensitive the computed Newton step is with respect to errors in the iteration of the unknown Fourier coefficients in Eq. \ref{eq:hbres}. If the local resonators are placed close to the nodes of mode shapes, which increases the condition number, convergence failures could happen in both harmonic balance and time-domain integration. 
Therefore, we use the linear solutions at the beginning frequency $\omega_s$ as the scaling matrix $\mathbf{\sigma}$:
\begin{equation}
\mathbf{\sigma}=\mathrm{diag} \left( \left|\left(-\omega_s^2 \mathbf{M}+\mathbf{K}\right)^{-1}\mathbf{F_{ex}} \right| \right).
\label{eq:scale}
\end{equation}
$\sigma$ is also known as the Jacobi preconditioner \cite{golub2013matrix}, the scaled unknowns are $\mathbf{\sigma}^{-1}\mathbf{u}$ with the similar order of magnitude, which reduces the condition number of the iteration problem and increases the convergence. 

An isotropic damping ratio $0.005$ is used for the host beam and local resonators to avoid numerical instability. The results of tip transmissibilities of the nonlinear metamaterial are shown in Fig. \ref{fig:fre}. The linear case is realized by prescribing a small base excitation acceleration $\ddot{w}_b$. The finite element method \cite{hu2018general} utilizes one-dimensional two-node Euler-Bernoulli beam elements. The element mass and stiffness matrices are modified to introduce the relative displacement $w_{rj}$ of the local resonators, thus enabling the inertia and reaction forces at the nodes of the resonators' positions. In Fig. \ref{fig:fre}(a), the transmissibility from the modal superposition method agrees with that from the FEM method, which means the modal superposition method proposed in Eq. \ref{eq:w} is sufficient. For the nonlinear case, the transmissibilities under different base excitations are solved by the harmonic balance method with harmonic order $H=11$. Since the modal weight $\eta_i$ not only contains the fundamental harmonic, the third and higher harmonic generations may occur. Thus, we take an example of the third harmonic generation at the free tip of the host beam. The ratio between the displacement amplitude of the third harmonic and that of the first-order harmonic is given as $Q_{31}$:
\begin{equation}
    Q_{31}=\left|\frac{\sum_{i=1}^N \hat{\eta}_i\left(3\right)\phi_i\left(L\right)}{\sum_{i=1}^N \hat{\eta}_i\left(1\right)\phi_i\left(L\right)}\right|.
    \label{eq:3rd}
\end{equation}
By varying the base excitation acceleration $\ddot{w}_b \in \left(0, 40g\right]$, the third harmonic generation ratio $Q_{31}$ is illustrated in Fig. \ref{fig:3rd}.  Besides the third harmonic generation at the linear resonant frequency $\omega_r$, the third harmonics also appear at the modal frequencies of the nonlinear metamaterial due to the coupling with nonlinear local resonators. This phenomenon points to the modal energy transfer from low to high frequency, which has been proven useful for modal response attenuation \cite{vakakis2008nonlinear}. With the increase of the base excitation, the ratio of the third harmonic becomes larger. However, the strongest third harmonic ratio is no more than 2.5\% under $\ddot{w}_b=40g$, which means the linear vibration modes still dominate the dynamics of the host beam. In Fig. \ref{fig:fre}, we only show the results of the fundamental harmonic induced bandgap since the generated harmonics are much higher than the region of interest near the linear resonant frequency $\omega_r$.

Similar to the dispersion analysis in Sec. \ref{subsec:dispersion}, the bandgap in Fig. \ref{fig:fre}(a) to (c) is gradually broadened due to the increase of nonlinear damping with the increase of excitation level. To maintain consistency, we modify the definition of the attenuation range of the bandgap with the actual length $L$ rather than the unit length of the host beam in Eq. \ref{eq:bandrange}. Regarding the bandgap range defined by $-20$ dB attenuation, the lower bound gradually shifts to a lower frequency and agrees  with the range of the dispersion analysis. For the bandgap's upper bound, there's a deviation between the result from frequency responses and that from dispersion analysis. This deviation is mainly caused by the number of the local resonators \cite{sugino2016mechanism} and the boundary conditions of the host beam. When the nonlinearity is weak, the solutions of both methods will converge by adding unit cells \cite{Zhao2023patra}. However, when nonlinearity is relatively strong, the interactions among nonlinear resonators themselves and with mode shapes of the host beam can not be neglected, which emphasizes the importance of frequency response analysis of a finite beam model. 
The spatial frequency analysis in  Fig. \ref{fig:fre}(d) to (f) shows the overall dynamic response of the metamaterial beam. It can be seen that the attenuation range starts from the first local resonator at position $x=0.14$m. As nonlinear damping increases, the bandgap becomes less marked \cite{wang2015wave}. 

\begin{figure}
\centering
\includegraphics[width=1\columnwidth,page=7]{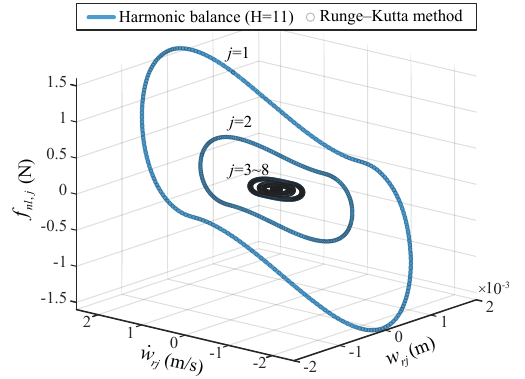}
\caption{Limit cycles of local resonators with their nonlinear forces at the linear resonant frequency $\omega_r$.}
\label{fig:cyc}
\end{figure}

Besides the bandgap, the harmonic balance method also yields the limit cycles of nonlinear local resonators at the fundamental resonant frequency $\omega_r$ under base excitation $\ddot{w}_b=20$g, shown in Fig. \ref{fig:cyc}. The results from the harmonic balance method synthesized from $H=11$ harmonic orders show good agreement with time domain integrations. From the $1^{st}$ to $8^{th}$ local resonator, the relative displacement and velocity amplitude decrease with the attenuation of the traveling flexural wave. With the increase of vibration amplitudes, the limit cycles indicate more nonlinearities due to the amplitude-dependent nonlinear damping forces $\mathbf{F_{nl}}$. Thus, the bandgap broadening effect near the resonant frequency $\omega_r$ is mainly due to the nonlinear damping effect of the resonators that are located near the clamped side of the host beam. 

\begin{figure}
\centering
\includegraphics[width=1\columnwidth,page=8]{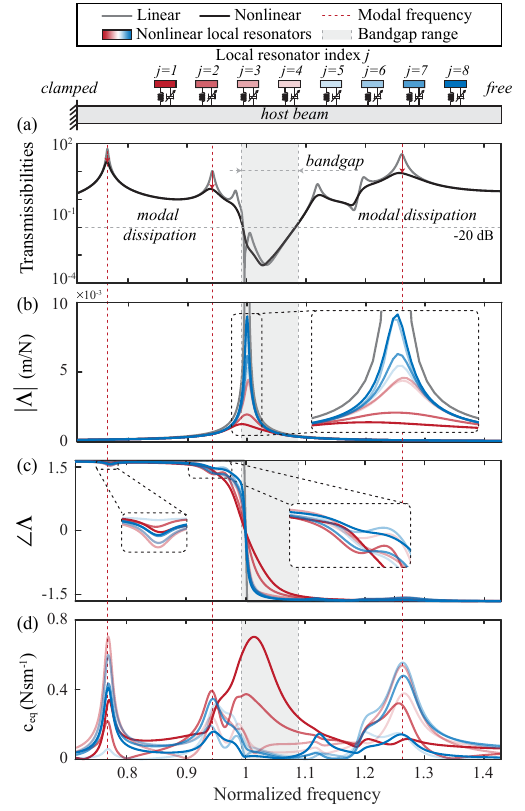}
\caption{Nonlinear modal dissipation in the metamaterial. (a) Vibration attenuation at modal frequencies; (b) and (c): Amplitudes and phases of frequency response functions of local resonators; (d) Equivalent damping of local resonators. }
\label{fig:nldamping}
\end{figure}

\subsection{Nonlinear Modal Dissipation}
\label{subsec:dissipation}
It is well-known that locally resonant metamaterials can create a bandgap for vibration mitigation of the host structure due to the spatial evanescent wave propagation starting from the resonance frequency of the local resonators \cite{liu2000locally}. On the other hand, nonlinear vibration absorbers have also been proven useful considering their broadband vibration mitigation ranges, which facilitate the redistribution of vibration energy over multiple vibration modes of the host structure and result in an efficient modal dissipative capacity of the host structure \cite{vakakis2008nonlinear}. In this part, we discuss the nonlinear modal dissipation in the proposed nonlinear metamaterial, which leads to the further broadening effect of the vibration attenuation range.

To investigate the nonlinear modal dissipation in the proposed metamaterial, we first look closely at the frequency response of the nonlinear local resonators. With the formulations in Sect. \ref{subsec:frf}, the fundamental harmonic \textit{Frequency Response Function} $\Lambda_j$ for each local resonator can be written as:
\begin{equation}
    \Lambda_j=\frac{W_{rj}}{\omega^2 m_j \left( W_b+W\right)}=\frac{1}{k_j-\omega^2m_j+\iu \omega c_{eq}},
\end{equation}
where $W_{rj}$ and $W$ are the fundamental harmonic amplitude of $w_{rj}$ and $w$. The definition of equivalent damping $c_{eq}$ in Eq. \ref{eq:ceq} can be applied here to demonstrate the effect of nonlinear damping from the local resonators with their fundamental harmonics calculated from the harmonic balance method. 

The frequency response of each nonlinear local resonator and their equivalent damping is shown in Fig. \ref{fig:nldamping} under $\ddot{w}_b=20$g. The bandgap formed around the resonant frequency of the local resonators gives the main vibration attenuation range of the nonlinear metamaterial. Within the bandgap range, the frequency responses of the nonlinear local resonators differ from those of the linear resonators, demonstrating the effect of amplitude-dependent nonlinear damping. With the right-going flexural wave, the nonlinear resonators close to the base excitation point present stronger equivalent damping $c_{eq}$ compared with those near the free tip of the host beam. This can be observed by the gradual transitions from the red curves with a larger $c_{eq}$ to the blue curves with a smaller $c_{eq}$ in Fig. \ref{fig:nldamping}(d). This nonlinear bandgap induced by nonlinear damping forms the first vibration attenuation range of the proposed nonlinear metamaterial due to the spatial decay of wave propagation and is also broadened by the nonlinear damping effect.

\begin{figure}
\centering
\includegraphics[width=1\columnwidth,page=9]{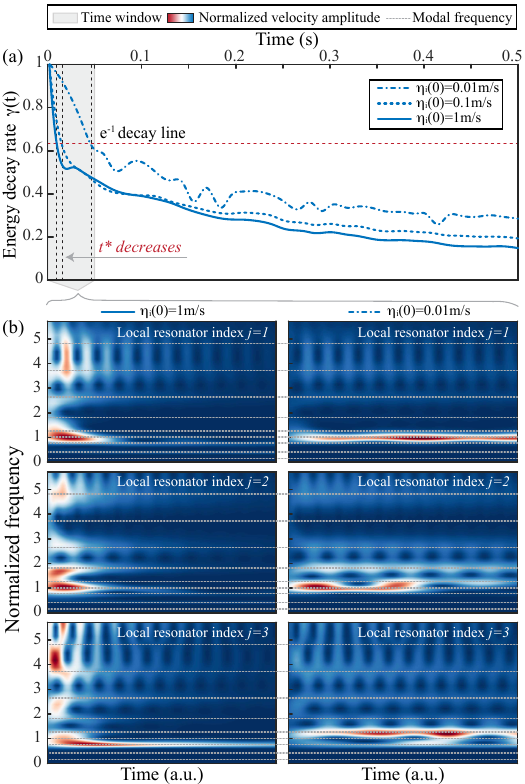}
\caption{The envelope of the energy decay rate and the wavelet transform of the nonlinear metamaterial after initial impact at $t=0$. (a) The envelope of the energy decay rate $\gamma(t)$ under three different impact velocities; (b) Wavelet transforms of the velocities $\dot{w}_{rj}$ of the first three nonlinear local resonators with the initial impact velocities $\eta_i(0)=1$m/s and $\eta_i(0)=0.01$m/s.}
\label{fig:transfer}
\end{figure}

Different from the bandgap range where the right propagating wave is eventually attenuated, the influence of reflections, in other words, the modes of the host beam, dominate the dynamical response of the metamaterial beam outside the bandgap range. These modes are not only coupled with each other, but they also interact with nonlinear resonators, which leads to nonlinear modal dissipation in this metamaterial beam. As shown in  Fig. \ref{fig:nldamping}(a), the transmissibilities at the modal frequencies indicated with red dash lines are much lower than its corresponding linear case, which means the mechanical energy is redistributed among different modal frequencies of the host beam and the nonlinear local resonators for efficient modal energy dissipation by different modes and nonlinear damping effect.  For a damped linear resonator, the damping effect flattens the slope change of its phase around its resonant frequency, as shown with the gray curve in Fig. \ref{fig:nldamping}(c). For a nonlinear local resonator attached to a finite-length beam, the equivalent damping $c_{eq}$ is determined by its velocity amplitude and the amplitudes of modal excitations. Therefore, there are multiple slope changes due to the modal frequency excitation from the movement of the host beam. As shown in Fig. \ref{fig:nldamping}(c), local slope changes of phase curves can be observed corresponding to modal frequencies. The increase of nonlinear damping around modal frequencies can also be observed  via $c_{eq}$ in Fig. \ref{fig:nldamping}(d). In particular, those resonators near the middle part and free end of the host beam contribute significantly to vibration attenuation at modal frequencies, which forms a contrast within the bandgap range that only the resonators near the vibration source are more effective. 
This nonlinear modal dissipation features a second mechanism besides the nonlinear bandgap and further broadens the range for vibration attenuation at modal frequencies.

In order to validate the efficient modal dissipation capability, we utilize the numerical time-domain integration in Eq. \ref{eq:td} with different modal velocities as initial conditions. The total kinetic energy $T(t)$ of the system can be represented with the kinetic energy of the host beam and that of the local resonators as:
\begin{equation}
T(t)=\frac{1}{2} \int_0^L \rho_0\left(\frac{\partial w}{\partial t}\right)^2 d x+\sum_{j=1}^S \frac{1}{2} m_j \dot{w}_{rj}^2.
  \label{eq:T}
\end{equation}
The total potential energy of the system $V(t)$ is represented by the strain energy of the host beam and the elastic energy of the local resonators:
\begin{equation}
V(t)= \frac{1}{2} \int_0^L  D_0\left(\frac{\partial^2 w}{\partial x^2}\right)^2 dx +\sum_{j=1}^S \frac{1}{2} k_j w_{rj}^2.
  \label{eq:V}
\end{equation}

Let $\gamma(t)$ denote the ratio of the instantaneous total energy $T(t)+V(t)$ to the input energy $T(t_0)+V(t_0)$ induced by initial condition:
\begin{equation}
  \gamma(t)=\frac{T(t)+V(t)}{T(t_0)+V(t_0)},
  \label{eq:decay}
\end{equation}
where $\gamma(t)$ describes how fast the initial energy is dissipated due to the linear and nonlinear damping effect in metamaterials. By following the definition of linear oscillators, we use $t^*$ to denote the time needed for the total energy drops by a factor of $e^{-1}$ of its initial value, which indicates the linear damping coefficient by a time inverse \cite{gzal2021extreme}. 


The results in Fig. \ref{fig:transfer}(a) show three initial modal velocities that mimic weak to strong impact conditions of the proposed nonlinear metamaterial.  The total energy of the three cases decays at different rates along the time evolution.  Under strong impact conditions, the whole nonlinear metamaterial's energy decays faster than in the other two cases, which indicates a stronger modal dissipation ability by the nonlinear damping effect.  Due to the local resonators, the modal frequencies of the metamaterial beam are altered. By solving the eigenvalue problem of the underlying linear system in Eq. \ref{eq:matrix}, the modal frequencies of the local resonant metamaterial beam are indicated with gray dash lines in Fig. \ref{fig:transfer}(b). Compared to the small initial impact case, the wavelet transforms of the strong impact case demonstrate stronger modal coupling with the modal frequencies. This modal coupling not only originates from the nondiagonal characteristic of the mass matrix for the underlying linear system in Eq. \ref{eq:MK}, but the nonlinear reaction forces further facilitate this coupling by mixing the states of the local resonators with the modal coordinates. For the strong impact case, more higher modal frequency components can be observed. This effect can be understood as low-frequency to high-frequency nonlinear energy transfer resulting for faster energy dissipation of the host system \cite{vakakis2008nonlinear}. The modal dissipation ability of the proposed nonlinear metamaterial helps the redistribution of the energy into nonlinear local resonators and higher vibration modes, which shows the potential application of this nonlinear metamaterial for shock wave attenuation. 

\section{Experiments}

The observed theoretical  bandgap broadening effect and nonlinear modal dissipation are verified experimentally. 
We first verify the nonlinear frequency response of the inertia amplifiers as local resonators for the metamaterial. Then, the nonlinear transmissibilities of the metamaterial are recorded under different base excitation conditions.  

    \subsection{Setup}

    \begin{table}[t!]
     \footnotesize
    \caption {Parameters in Experiment} \label{para}
    \begin{center}
    \setlength{\tabcolsep}{2mm}{
    \begin{tabular}{llll}
    \toprule
    \multicolumn{4}{c}{\textbf{Nonlinear Damping induced Metamaterial}}     \\
    \midrule
    \multicolumn{4}{l}{\textbf{Host Beam}}                                \\
    Size   & 600$\times$40$\times$12 (mm$^3$) & Material  & Versatile Plastic \\
     Density &  980 (Kg/m$^3$) & $E$  & 1.53 (GPa)
    \vspace{0.05in} \\
    \multicolumn{4}{l}{\textbf{Nonlinear Local Resonator}}                \\
     Material  & Versatile Plastic & $R$  & 20 (mm)  \\
     $d$  & 40 (mm)  & $\theta$ & $\ang{35}$  \\
     $m_0$   &  3.53 (g)   & $k_0$ & 15.5 (kN/m) \\ $\omega_r$ & 2$\pi\times$210.7 (rad/s) & \multicolumn{2}{l}{$x_j=140:60:560$  (mm) }  \\
    \bottomrule
    \end{tabular}}
    \end{center}
    \end{table}

    \begin{figure}
    \centering
    \includegraphics[width=1\columnwidth,page=10]{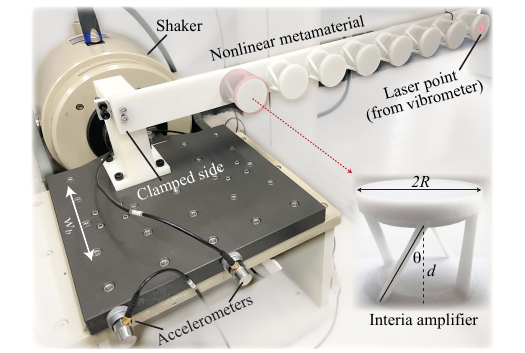}
    \caption{Experimental setup. }
    \label{fig:setup}
    \end{figure}

    \begin{figure}
    \centering
    \includegraphics[width=1\columnwidth,page=11]{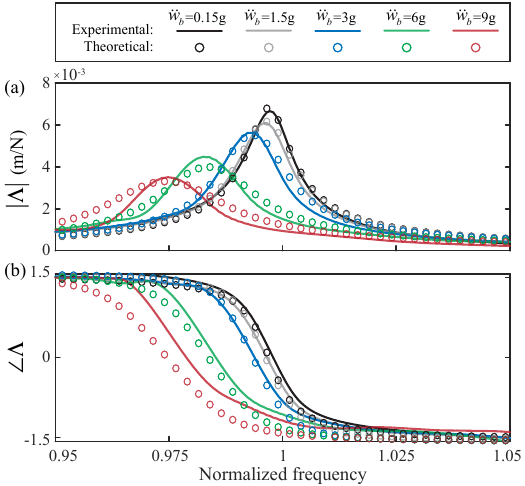}
    \caption{Experimental and theoretical frequency response function $\Lambda$ of an inertia amplifier with $\theta=\ang{35}$ under different base excitation accelerations $\ddot{w}_b$.}
    \label{fig:exp_res}
    \end{figure}

    \begin{figure*}
    \centering
    \includegraphics[width=2\columnwidth,page=12]{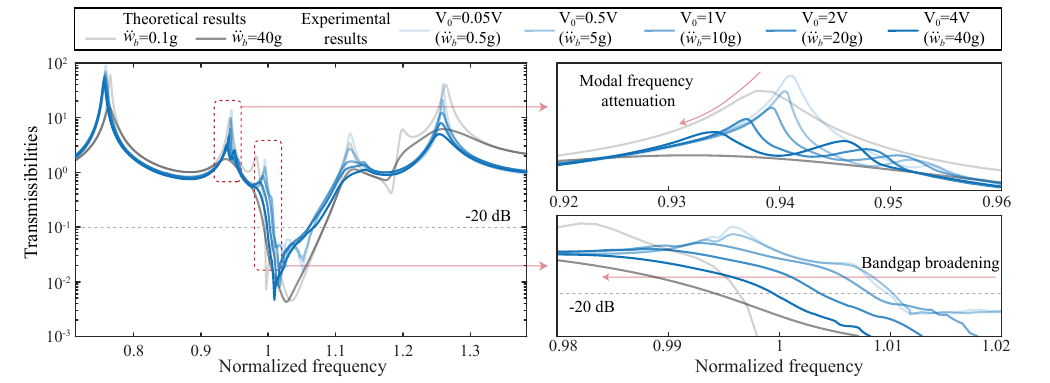}
    \caption{Experimental and theoretical transmissibilities of the nonlinear metamaterial under different base excitation voltages $V_0$ and accelerations $\ddot{w}_b$. }
    \label{fig:exp_nm}
    \end{figure*}

Figure \ref{fig:setup} shows the experimental setup. The nonlinear metamaterial is made of versatile plastic and consists of a 0.6-meter host beam and a series of nonlinear local resonators spaced 60 mm apart. The prototype is printed with Selective Laser Sintering 3D printing (EOSINT P760) in one piece. The parameters of the nonlinear metamaterial are shown in Tab.~\ref{para}. To maintain stability during excitation, the nonlinear metamaterial is then hung vertically. One end of the metamaterial is clamped to a shaker (VE-5120) for base excitation, resulting in a clamped-free boundary condition. A Polytec laser Doppler vibrometer (LDV) is used to record the out-of-plane velocity field at any point along the structure, such as the clamped and free end of the host beam. The data is acquired in the time domain through repeated acquisitions.

\subsection{Experimental Results}

Before the experiments on the nonlinear metamaterial shown in Fig. \ref{fig:setup}, we first validate the nonlinear damping effect of an individual inertia amplifier. As shown in the enlarged view in Fig. \ref{fig:setup}, we chose a tilted angle of $\ang{35}$. The 3D-printed inertia amplifier is fixed to the shaker. In order to observe the nonlinear damping effect, a slow sweep signal with a rate of 0.1 Hz/s ranging from 180 Hz to 230 Hz is applied to the base of the nonlinear local resonator at a constant excitation force. Time domain velocity responses at different points on the top disk and excitation base of the resonator are recorded with the LDV. 

The resonator's experimental and theoretical nonlinear frequency response is illustrated in Fig. \ref{fig:exp_res} with different base accelerations from 0.15 g to 9 g. The theoretical results are obtained using Eq. \ref{eq:eom_res} by the harmonic balance method mentioned in Sec. \ref{subsec:frf}, which show good agreement with experimental results. It can be seen that the resonant frequency of the resonator slightly shifts to a lower frequency with the increase of excitation force, which indicates an increase of the nonlinear damping effect studied previously, and a stiffness reduction that has also been observed in \cite{van2021inherent}. Therefore, we employed a cubic fitting of the stiffness versus the displacement amplitude $W_r$ of the resonator, which returns the linear stiffness $k_0$ when $W_r$ is small. A linear damping ratio $\zeta_0=0.005$ is also applied here.  The nonlinear damping effect becomes prominent under large base accelerations, reducing the amplitude of its frequency response function and broadening its bandwidth. This amplitude-dependent nonlinear damping can also be observed from the phase angle of the frequency response Fig. \ref{fig:exp_res}(b). The slope of the phase transition around resonances is smaller under large excitation, which qualifies the theoretical model discussed in Sec. \ref{subsec:dispersion}.

Once the nonlinear damping effect on the single resonator level has been identified, the transmissibilities of the nonlinear metamaterial, combining the base beam and sequential resonators, are shown in Fig. \ref{fig:exp_nm}. We excite the platform where the clamped side of the host beam is mounted at constant excitation force (controlled by the excitation voltage $V_0$ of the amplifier). As in the previous experiment, sweep signals slowly varying from 100 Hz to 350 Hz at 0.1 Hz/s are applied. 

Due to the influence of modal responses on the excitation base, the experimental transmissibility of the host beam is calculated as the tip velocity of the host beam normalized by the input acceleration. It is further nondimensionalized with the excitation frequency. The theoretical transmissibility takes the same form as explained in Sec. \ref{subsec:frf}. We use two theoretical cases ($\ddot{w}_b=0.1$ g and $\ddot{w}_b=40$ g) to indicate the effective range of the nonlinear metamaterial and compare these with the experimental results. It can be seen that the general trend and bandgap range of the experimental and theoretical results agree, which validates the analyses in Sec. \ref{sec:theor}. 
    
By closely checking the enlarged views enclosed in red dashed squares, we can observe the bandgap broadening effect and the nonlinear energy transfer at modal frequencies. Under large excitation levels, the bandgap is broadened not only due to the nonlinear damping effect, but also the shifting of the stiffness of nonlinear local resonators. It should be noted that the bandgap degenerating effect is not obvious since the selected $\theta$ in the experiments is not small. Furthermore, the modal frequency peak around normalized frequency 0.9 gradually splits into two smaller peaks with the increase of the excitation level, highlighting the effect of the nonlinear damping. 

    \begin{figure*}
    \centering
    \includegraphics[width=2\columnwidth,page=13]{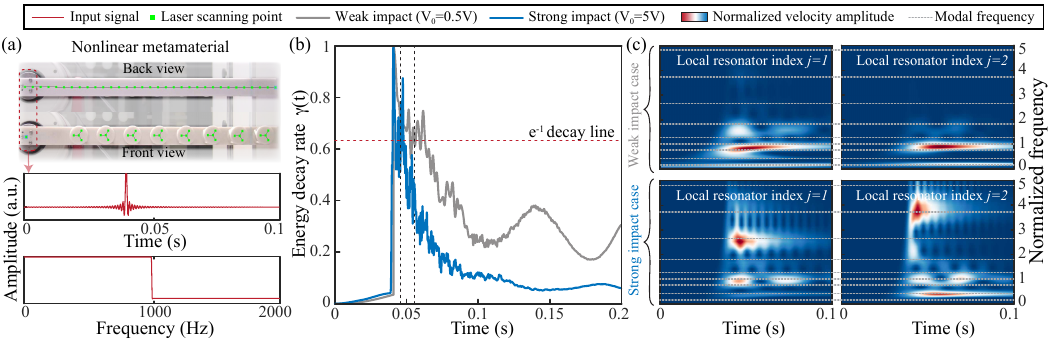}
    \caption{Experimental shock wave attenuation of the nonlinear metamaterial under different impact voltages $V_0$. (a) Experimental setup of the nonlinear metamaterial with the laser scanning points and the profile of the input impact signal; (b) Experimental energy decay rates of the nonlinear metamaterial under different impact intensities; (c) Experimental wavelet transforms of the velocities $\dot{w}_{rj}$ of the first two nonlinear local resonators for the two different impact cases.}
    \label{fig:exp_impact}
    \end{figure*}

To further demonstrate the nonlinear modal coupling and dissipation ability of the nonlinear metamaterial beyond its bandgap range, experimental impacts have been applied at the clamped side of the nonlinear metamaterial. The experimental setup is shown in Fig. \ref{fig:exp_impact}(a), multiple laser scanning points are defined at the front and back sides of the nonlinear metamaterial to measure the velocity responses of the nonlinear local resonators and the host beam under different impact intensities. For each measurement of the out-of-plane velocity, a repeatable raised cosine pulse signal, with a bandwidth for modal frequencies lower than 1000 Hz, is excited by the shaker at the clamped side of the nonlinear metamaterial. The scanning points distributed along the host beam are spaced 20 mm apart. 

The time evolution of the energy decay rates $\gamma(t)$ under different impact intensities are calculated with Eq. \ref{eq:T} to Eq. \ref{eq:decay}, in which the displacement and its derivative versus beam length are acquired by numerical integration of the velocity response and finite difference method, respectively. In Fig. \ref{fig:exp_impact}(b), a stronger impact not only leads to faster energy dissipation of the system but also suggests energy redistribution to high frequency modes of the nonlinear metamaterial, which agrees with the modal dissipation ability studied in Sec. \ref{subsec:dissipation}. By closely checking the wavelet transforms of the first two nonlinear local resonators in Fig. \ref{fig:exp_impact}(c), there exist more frequency components correspond to higher modal frequencies for the strong impact case, which indicates the modal coupling ability of the nonlinear metamaterial for efficient energy redistribution of the host beam for shock wave and impact attenuation.
    
\section{Conclusion}
This paper presents a practical, tunable nonlinear resonator based on the nonlinear damping effect induced by rotational inertia amplifiers. This resonator is used to create a nonlinear metamaterial for broadband vibration attenuation, combining a broader bandgap and general modal vibration dissipation within the host structure. Revisiting the nonlinear damping effect mechanism, we establish the nonlinear dispersion relationships for a semi-infinite nonlinear metamaterial case. More importantly, with respect to practical applications, the nonlinear frequency response of a finite structure is studied with modal analysis and Alternating Frequency Time (AFT) multiple harmonic balance methods for general nonlinearities in nonlinear metamaterials. 
The theoretical results reveal that the bandgap is broadened  with the increase of excitation level. Especially, the nonlinear interactions between the local resonators and the mode shapes of the host beam lead to efficient modal frequency dissipation ability in the proposed metamaterial. Finally, experiments were carried out with both the single nonlinear resonator level and the full metamaterial system. The experimental results validate both the nonlinear bandgap and modal dissipation as mechanisms for broadband and shock wave attenuation. By incorporating the effect of nonlinearity with the concept of conventional locally resonant metamaterials, our findings enable new possibilities for vibration attenuation beyond the conventional linear bandgap range.

\appendix
\section{Modal Superposition}

The mass-normalized shape function of the host beam with clamped-free boundary condition can be shown as:
\begin{equation}
\begin{aligned}
&\phi_{i}(x)=\frac{1}{\sqrt{\rho_0 L}}\left[\cos \left(\frac{\lambda_{i} x}{L}\right)-\cosh \left(\frac{\lambda_{i} x}{L}\right)\right. \\
&\left.\quad+\left(\frac{\sin \lambda_{i}-\sinh \lambda_{i}}{\cos \lambda_{i}+\cosh \lambda_{i}}\right)\left(\sin \left(\frac{\lambda_{i} x}{L}\right)-\sinh \left(\frac{\lambda_{i} x}{L}\right)\right)\right], \\
&i=1,2, \ldots, N
\end{aligned}
\label{eq:mode}
\end{equation}
where $\lambda_{i}$ is the positive eigenvalue of the characteristic equation read as:
\begin{equation}
\cos\lambda_{i}\cosh\lambda_{i}+1=0.
\label{eq:11}
\end{equation}

By substituting Eq. \ref{eq:w} into Eq. \ref{eq:govern}, applying the orthogonality conditions with $\phi_s(x)$, and integrating over the span of the host beam, Eq. \ref{eq:govern} is transformed into:
\begin{equation}
\begin{aligned}
&\ddot{\eta}_i+\omega_i^2 \eta_i-\sum_{j=1}^{S} \left(k_{j} w_{rj}+ {c_j}|{\dot w}_{rj}|{\dot w}_{rj}\right) \phi_i(x_j) \\
&=-\rho_0 \ddot{w}_{b}\int_{x=0}^{x=L}\phi_i(x)dx,
\end{aligned}
\label{eq:trans1}
\end{equation}
where $\omega_i$ is the natural frequency of the $i$-th mode of the host beam. Follow the same procedure to substitute Eq. \ref{eq:w} into Eq. \ref{eq:eom_res2}, it gives:
\begin{equation}
\begin{aligned}
    &{m_j}\left(\sum_{i=1}^N \ddot{\eta}_i\phi_i(x_j)+{\ddot w}_{rj}\right) + {c_j}|{\dot w}_{rj}|{\dot w}_{rj} \\ & + {k_j}w_{rj} =  - {m_j}{{\ddot w}_b}.
\end{aligned}
\label{eq:trans2}
\end{equation}

The nonlinear reaction forces induced by the local resonators that are applied on the host beam can be represented by Eq. \ref{eq:trans2}. Therefore, Eq. \ref{eq:trans1} can be rewrite as:
\begin{equation}
\begin{aligned}
&\ddot{\eta}_{i}+\omega_{i}^{2} \eta_{i}+\sum_{j=1}^{S} m_{j} \phi_{i}\left(x_{j}\right) \sum_{i=1}^{N} \ddot{\eta}_{i} \phi_{i}\left(x_{j}\right) \\
&+\sum_{j=1}^{S} m_{j} \ddot{w}_{rj} \phi_{i}\left(x_{j}\right)=q_{i}, \quad r=1,2, \ldots, N
\end{aligned}
\label{eq:trans3}
\end{equation}
where
\begin{equation}
q_{i}=-\ddot{w}_{b}\left(\int_{x=0}^{x=L} \rho_0 \phi_{i}(x) \mathrm{d} x+\sum_{j=1}^{S} m_{j} \phi_{i}\left(x_{j}\right)\right).
\label{eq:q}
\end{equation}

By combining Eq. \ref{eq:trans2} and Eq. \ref{eq:trans3}, the matrix form equations of the nonlinear metamaterial can be achieved.

\printcredits

\section*{Acknowledgment}

The authors acknowledge support from the ETH Research Grant (ETH-02 20-1), the H2020 FET-proactive project METAVEH under the grant agreement 952039, and the SNSF R'Equip grant 206021\textunderscore205418.

\bibliographystyle{model1-num-names}

\bibliography{ref}




\end{document}